\documentstyle[12pt]{article}

\begin{document}

\newcommand{\dfrac}[2]{\displaystyle{\frac{#1}{#2}}}

{\it University of Shizuoka}

\hspace*{9.5cm} {\bf US-96-10 }\\[-.3in]

\hspace*{9.5cm} {\bf AMU-96-11}\\[-.3in]

\hspace*{9.5cm} {\bf November 1996}\\[-.3in]


\vspace*{.1in}

\vspace*{.5in}

\begin{center}

{\large\bf Analytical Expressions of Masses 
and  Mixings  }\\[.2in]

{\large\bf in a Democratic Seesaw Mass Matrix Model}\\[.5in]

{\bf Yoshio Koide}\footnote{
E-mail: koide@u-shizuoka-ken.ac.jp} \\

Department of Physics, University of Shizuoka \\ 
395 Yada, Shizuoka 422, Japan \\[.1in]

and \\[.1in]

{\bf Hideo Fusaoka}\footnote{
E-mail: fusaoka@amugw.aichi-med-u.ac.jp} \\

Department of Physics, Aichi Medical University \\ 
Nagakute, Aichi 480-11, Japan \\[.5in]

{\large\bf Abstract}\\[.1in]

\end{center}

\begin{quotation}
On the basis of a seesaw-type mass matrix model 
$M_f\simeq m_L M_F^{-1} m_R$  for quarks and leptons $f$, 
analytical expressions of the masses and mixings of the fermions 
$f$ are investigated.  
Here, the matrices $m_L$ and $m_R$ are common to all $f$ 
(up- and down-; quarks and leptons), and 
the matrix $M_F$ characterizing the heavy fermion sector
has the form [(unit matrix)+ (democratic-type matrix)].
An application to the quark sectors is discussed. 
\end{quotation}

\newpage
\centerline{\bf \S 1. Introduction}

\vglue.05in

Why is the top quark mass $m_t$ so enhanced compared with the 
bottom quark mass $m_b$? Why is the $u$-quark mass $m_u$ of 
the order of the $d$-quark mass $m_d$\,?
In most models, in order to understand $m_t\gg m_b$, 
it is inevitable to bring in a parameter which takes hierarchically 
different values between up- and down-quark sectors.
However, from the point of view of the ``democracy of 
families", such a hierarchical difference seems to be unnatural.
What is of great interest to us is 
whether we can find a model in which $M_u$ and $M_d$ are almost 
symmetric in their matrix structures and in their 
parameter values.

Recently, by applying the  so-called ``seesaw" mechanism${}^{1)}$ 
to quark mass matrix,$^{2)}$ the authors${}^{3)}$ have proposed a model
which provides explanations of both
$m_t \gg m_b$ and $m_u \sim m_d$, while keeping the model 
``almost" up-down symmetric. 
The essential idea is as follows: 
the mass matrices $M_f$ of quarks and leptons $f_i$ 
($i=1,2,3$: family index) are given by
$$
M_f \simeq -m_L M_F^{-1} m_R, \eqno(1.1)
$$
where $F_i$ denote heavy fermions $U_i$, $D_i$, 
$N_i$ and $E_i$, corresponding to $f_i = u_i, d_i, \nu_i$ 
and $e_i$, respectively. They have assumed that the mass 
matrix $m_L$ ($m_R$) between $f_L$ $(f_R)$ and $F_R$  $(F_L)$ 
is common to all $f = u, d, \nu, e$ 
(i.e., independently of up-/down- and quark-/lepton- sectors) 
and $m_R$ is proportional to $m_L$, i.e., $m_R=\kappa m_L$.
The variety of 
$M_f$  ($f = u, d, \nu, e)$ comes only from the variety 
of the heavy fermion matrix $M_F$  $(F = U, D, N, E)$. 
If we take a parametrization which gives det$M_U\simeq 0$ 
in the up-quark sector, but which does not give det$M_D\simeq 0$
in down-quark sector, 
the model can provide $m_t \gg m_b$, keeping the model 
``almost" up-down symmetric 
because of the factor $M_F^{-1}$ in the seesaw expression (1.1).
On the other hand, they have taken  $M_F=m_0 \lambda O_f$  as 
the form of the heavy fermion mass matrix $M_F$, where
$$
 O_f = \left(\begin{array}{ccc}
1 & 0 & 0 \\
0 & 1 & 0 \\
0 & 0 & 1 \end{array} \right) + b_f e^{i\beta_f} \left(
\begin{array}{ccc}
1 & 1 & 1 \\
1 & 1 & 1 \\
1 & 1 & 1 \end{array} \right) 
\equiv {\bf 1} + 3b_f e^{i\beta_f} X 
\ , \eqno(1.2)
$$
and $\lambda$ is an enhancement factor with $\lambda\gg \kappa \gg 1$.
Note that the inverse of the matrix $O_f$ is again given by 
the form${}^{4)}$ [(a unit matrix) + (a democratic matrix)], i.e., 
$$
O_f^{-1} = {\bf 1} + 3a_f e^{i\alpha_f} X \ , \eqno(1.3) 
$$
with
$$
a_f e^{i \alpha_f} = -\dfrac{b_f e^{i\beta_f}}{1 + 3b_f e^{i\beta_f}} \ . 
\eqno(1.4)
$$
Thus, we can provide top-quark mass enhancement $m_t \gg m_b$ 
in the limit of $b_u e^{i \beta_u} \rightarrow -1/3$, 
because it leads to $|a_u| \rightarrow \infty$. On the 
other hand, since a democratic mass matrix${}^{5)}$ makes only one 
family heavy, we can keep $m_u \sim m_d$. 

They have taken 
$$
m_L = \dfrac{1}{\kappa} m_R = m_0 Z \equiv
 m_0 \left( \begin{array}{ccc}
z_1 & 0 & 0 \\
0 & z_2 & 0 \\
0 & 0 & z_3 \end{array} \right) \ , \eqno(1.5)
$$
where $z_i$ are normalized as $z_1^2+z_2^2+z_3^2=1$ and 
given by 
$$
\dfrac{z_1}{\sqrt{m_e}}=\dfrac{z_2}{\sqrt{m_\mu}}=
\dfrac{z_3}{\sqrt{m_\tau}}=\dfrac{1}{\sqrt{m_e+m_\mu+m_\tau}} \ , 
\eqno(1.6) 
$$
in order to give  the charged lepton mass matrix $M_e$ for
the case $b_e=0$,  i.e., $M_e = m_0 (\kappa/\lambda) Z^2$. 
They have obtained${}^{3)}$ reasonable quark mass ratios and 
Cabibbo-Kobayashi-Maskawa${}^{6)}$ (CKM) matrix parameters by taking 
$\kappa/\lambda=0.02$, 
$b_u=-1/3$, $\beta_u=0$, $b_d\simeq -1$ and $|\beta_d|\simeq 18^\circ$.

However although they numerically evaluated the behavior of the CKM 
matrix elements to the parameters $\kappa/\lambda$, $b_f$ and $\beta_f$, 
they did not give analytical expressions of the CKM matrix elements.
Therefore, of their results, we cannot see which are results only for 
a special choice of the parameters and which are (almost) 
parameter-independent ones.
For example,  they predicted a value  $|V_{cb}| = 0.0598$, 
which is somewhat 
large compared with the recent experimental value${}^{7)}$ 
$|V_{cb}| = 0.041 \pm 0.003$.
However, we cannot see whether the discrepancy is a fatal defect 
in this model or not.

What is of great interest to us is to clarify the general features of
the democratic seesaw mass matrix, without confining ourselves 
to the phenomenology of the quark masses and CKM matrix elements. 
It is also interesting to apply the model to other fermion systems, 
for example, to neutrino sector, a hypothetical fermion system, and so on.
For this purpose, it is inevitable to obtain analytic expressions of
the fermion masses $m_i^f$ and the family-mixing matrix $U_L^f$ 
for arbitrary values of the parameters $b_f$ and $\beta_f$, 
and not to give such the numerical study as in Ref.~3). 
In \S 3, we will give general expressions of the fermion masses $m_i^f$ 
for arbitrary $b_f$ and $\beta_f$, although the cases of $b_f=-1/3$ and 
$b_f\simeq -1$ have already been in Ref.~3).
In \S 4, we will obtain a general expression of the $3 \times 3$ 
family-mixing matrix $U^f_L$ for arbitrary values of the parameters 
$b_f$ and $\beta_f$.

Since the previous paper${}^{3)}$ put stress on the ``economy of adjustable 
parameters" 
of the model, the predictions were done by adjusting 
only three parameters $\kappa/\lambda$, $b_d$ and $\beta_d$. 
As a result, some of the predictions were in poor agreement 
with experiment.
In the present paper, we will loosen the parameter constraints in the 
previous model${}^{3)}$ (we will bring two additional phase parameters 
into the model).
As a result, the predictability of the model decreases. 
However, the purpose of the present paper is not to improve  
the previous quark mass matrix model, but to investigate more general 
features of a democratic seesaw mass matrix model 
without confining ourselves 
in the quark mass matrix phenomenology.

As an application of our general study to the quark sectors, 
in \S 5, we will discuss analytical expression of the CKM matrix.
In \S 6, we will give re-fitting of the CKM matrix parameters. 
Also, a possible shape of the unitary triangle 
$V_{ud} V_{ub}^{\ast} + V_{cd} V_{cb}^{\ast} 
+ V_{td} V_{tb}^{\ast} = 0$ 
in our model will be discussed. 

The final section \S 7 will be devoted to the summary and discussion.

\vglue.2in
\centerline{\bf \S 2. Assumptions for the model}

\vglue.05in

In the present model, quarks and leptons $f_i$ belong to 
$f_L = (2,1)$ and 
$f_R = (1, 2)$ of SU(2)$_L \times $SU(2)$_R$ 
and heavy fermions $F_i$ are vector-like, i.e., 
$F_L = (1, 1)$ and $F_R = (1, 1)$. 
The vector-like fermions $F=(F_1,F_2,F_3)$ acquire masses $M_F$ 
at a large energy scale $\mu=m_0 \lambda$.
The SU(2)$_L$ and SU(2)$_R$ symmetries are broken by 
Higgs bosons $\phi_L=(\phi_L^+, \phi_L^0)$ and 
$\phi_R=(\phi_R^+, \phi_R^0)$, which belong to (2,1) and (1,2) of
SU(2)$_L\times$SU(2)$_R$, at energy scales $\mu=m_0$ and 
$\mu=m_0 \kappa$, respectively.

Let us summarize the fundamental assumptions in the previous 
paper${}^{3)}$ 
before starting our analytical study of the democratic seesaw 
mass matrix model.

[Assumption I] \ \ The $6\times 6$ mass matrix $M$ for the fermions 
$(f, F)$ has a would-be ^^ ^^ seesaw" form 
$$
\left(\overline{f} \ \overline{F}\right)_L M 
\left(\begin{array}{c}
f \\ 
F \\ 
\end{array}\right)_R = \left(\overline{f} \ \overline{F}\right)_L 
\left(\begin{array}{cc}
0 & m_L \\ 
m_R & M_F \\ 
\end{array}\right) \left(\begin{array}{c}
f \\ 
F \\ 
\end{array}\right)_R \ \ , 
\eqno(2.1)
$$
$i.e.$, there is no Higgs boson which belongs to (2,2) of SU(2)$_L$ $\times$ 
SU(2)$_R$, in contrast with the conventional SU(2)$_L$ $\times$ SU(2)$_R$ 
model.$^{8)}$ 

[Assumption II] \ \ The structure of $m_R$ is the same as that of 
$m_L$ except for a constant coefficient $\kappa$, $i.e.$, 
$$
m_R = \kappa m_L  \ \ . 
\eqno(2.2)
$$

[Assumption III] \ \ The heavy fermion mass matrix $M_F$ takes a form 
[(a unit matrix) $+$ (a rank-one matrix)], $i.e.$, 
$$
M_F = m_0 \lambda_F ({\bf 1} + 3 b_f e^{i\beta_f} R_1) \ \ , 
\eqno(2.3)
$$
where $R_1$ is an arbitrary rank-one matrix.

The requirement that the matrix $R_1$ is a rank-one matrix is indispensable 
to realize that the choice det$M_F(b_f)=0$ makes a mass of only one 
fermion heavy, $i.e.$, $m_t \gg m_c > m_u$ with keeping $m_u \sim m_d$. 
Note that at this stage, it is not necessary to assume 
that the matrix $R_1$ has a democratic form as defined by (1.2). 
Without losing generality, we can take a favorite family-basis of the heavy 
fermions $F=(F_1, F_2, F_3)$. 
However, in order to obtain the successful fitting of the quark masses 
and CKM mixings in Ref.~3), the following assumption is essential. 

[Assumption IV] \ \ When we choose the family-basis where $R_1$ takes the 
democratic form 
$$
R_1=X \equiv  \frac{1}{3}\left(\begin{array}{ccc}
1 & 1 & 1 \\ 
1 & 1 & 1 \\ 
1 & 1 & 1 \\ 
\end{array} \right) \ \ , 
\eqno(2.4)
$$
the matrix $m_L$ takes a diagonal form $m_0 Z$, (1.5).

If we take another family-basis $(f', F') = (Af, AF)$, the mass matrix $M'$ 
for $(f', F')$ is given by 
$$
M' = \left(\begin{array}{cc}
0 & m'_L \\ 
m'_R & M'_F \\ 
\end{array}\right) = \left(\begin{array}{cc}
0 & Am_L A^\dagger \\ 
A m_R A^\dagger & A M_F A^\dagger \\ 
\end{array}\right) \ \ . 
\eqno(2.5)
$$
Without losing generality, we can choose a basis on which $M'_F$ takes 
a diagonal form 
$$
M'_F = m_0 \lambda_F\, {\rm diag} (1, 1, 1+3b_f e^{i\beta_f}) \ \ . 
\eqno(2.6)
$$ 
Therefore, Assumption IV can be replaced with the following expression:

[Assumption IV$'$] \ \ On the family-basis on which $M'_F$ is diagonal, 
the mass matrices $m'_L$ and $m'_R$ are given on the family-basis 
$f' = (f'_1, f'_2, f'_3)$ which consists of representations of 
the permutation group $S_3$ of three elements, $i.e.$, 
$$
\begin{array}{l}
f'_1 = \frac{1}{\sqrt{2}}(f_1 -f_2) \ \ , \\
f'_2 = \frac{1}{\sqrt{6}}(f_1 + f_2 - 2f_3) \ \ , \\
f'_3 = \frac{1}{\sqrt{3}}(f_1 + f_2 + f_3) \ \ , \\ 
\end{array} 
\eqno(2.7)
$$
where $f_i$ are fermion states in which $m'_L$ and $m'_R$ are 
diagonalized. 

In any expressions IV and IV$'$, it is essential that $M_F$ is given by a 
form [(a unit matrix ) $+$ (a democratic matrix)] on the family-basis 
on which $m_L$ and $m_R$ take diagonal forms. 
For a mechanism which generates such a democratic mass matrix, 
some ideas have been proposed:
a permutation symmetry of three elements $S_3$,$^{9)}$
a composite model based on an analogy of hadronic 
$\pi^0$-$\eta$-$\eta'$ mixing,$^{10)}$
a BCS-like mechanism,$^{11)}$  and so on. 
However, the purpose of the present paper is not to investigates 
the origin of the democratic mass matrix form. 
We do not touch the origin of the form (2.4).

In the numerical study for the quark sectors, the coefficient 
$\lambda_F$ will be assumed as $\lambda_U \simeq \lambda_D \equiv 
\lambda_Q \neq \lambda_E$, because the evolution effects 
of Yukawa coupling 
constants can be different according as the fermions have color or not, 
even if $\lambda_U=\lambda_D=\lambda_E$ at a unification energy scale. 

As we stated in \S 1, in the present paper, 
we will loosen  parameter constraints in the previous model 
and we will bring two additional phase parameters 
$\delta_2$ and $\delta_3$ into the CKM-matrix phenomenology.
We assume that the Higgs bosons $\phi_L$ and $\phi_R$ 
couple to the fermions universally, but with the degree of freedom of 
their phases, as follows:
$$
H_{Yukawa}=\sum_{i=1}^3 (\overline{u}_{L i} \ \overline{d}_{L i})
\left(y_{Li} \exp(i\delta^d_{Li})\right) \left(
\begin{array}{c}
\phi_L^+ \\
\phi_L^0 
\end{array}
\right) D_{Ri} $$
$$+  
\sum_{i=1}^3 (\overline{u}_{L i} \ \overline{d}_{L i})
\left(y_{Li} \exp(i\delta^u_{Li})\right) \left(
\begin{array}{c}
\overline{\phi}_L^0 \\
-\phi_L^- 
\end{array}
\right) U_{Ri} 
$$
$$
+ h.c. +(L\leftrightarrow R) + [(u,d,U,D)\rightarrow (\nu,e,N,E)]
\ , \eqno(2.8)   
$$
where $y_{Li}$ and $y_{Ri}$ are real parameters, and they are 
universal for the quark and lepton sectors.
Therefore, the matrices $m_L$ and $m_R$ in (2.1) are replaced with 
$$
m_L^f = m_0 P_L^f Z \equiv m_0\, {\rm diag}\left(z_1 
{\rm exp}(i \delta_{L_1}^f), 
z_2 {\rm exp}(i \delta_{L_2}^f), z_3 {\rm exp}(i \delta_{L_3}^f)\right) \ \ , 
\eqno(2.9)
$$
and $m_R^f = m_0 \kappa P_R^f Z$, respectively, where $P_L^f$ and 
$P_R^f$ are phase matrices. 

For these phase parameters, 
the CKM matrix is dependent only on
$$
P_{L}^{u\dagger} P_L^{d}\equiv P={\rm diag}(e^{i\delta_1}, 
e^{i\delta_2}, e^{i\delta_3}) \ .
\eqno(2.10)
$$
Of the three parameters $\delta_i$ ($i=1,2,3$), only two are observable.
Without losing generality, we can put $\delta_1=0$. 
In the present model, the nine observable quantities 
(five  quark mass  ratios and four CKM matrix parameters) are described by 
the seven parameters ($\kappa/\lambda$, 
$b_u$, $b_d$, $\beta_u$, $\beta_d$, $\delta_2$, $\delta_3$). 
Since we put the ansatz  ``maximal top-quark-mass enhancement" 
according to the Ref.~3), we fix $b_u$ and $\beta_u$ at $b_u=-1/3$ and 
$\beta_u=0$.
However, we still possess five free parameters.
In order to economize in the number of the free parameters, we will give 
some speculation on these parameters in the final section.
On the other hand, since the phases $\delta_{L i}^e$ and 
$\delta_{R i}^e$ are not observable,
we can put $P_L^e=P_R^e={\bf 1}$.

\vglue.2in
\centerline{\bf \S 3. General expressions of the fermion masses}

\vglue.05in

The general case $P_L^f \neq {\bf 1}$  does not change the previous
results${}^{3)}$ as far as the mass ratios are concerned.
Quark masses in terms of charged lepton masses have already been given 
in Ref.~3).
However, the previous expressions were only those for the cases of 
$(b_f=-1/3, \beta_f=0)$, and $(b_f\simeq -1, \beta_f\simeq 0)$. 
In the present paper, we will give general expressions for arbitrary 
values of $b_f$ and $\beta_f$.

Note that for the case 
$b_f = -1/3$ the seesaw expression (1.1) is not valid any longer
because of det$M_F=0$. 
In Fig.~1, we illustrate the numerical behavior of fermion masses 
$m_i^f$ ($i=1,2,\cdots ,  6$) versus the parameter $b_f$ 
which has been evaluated from the 
$6 \times 6 $ matrix (2.1)  without approximation (the behavior of 
$m_i^f$ with $i=1,2,3$ has been illustrated in Ref.~3).
As seen in Fig.~1, the third fermion is sharply 
enhanced at $b_f=-1/3$ for $\beta_f=0$. 
The calculation for the case $b_f\simeq -1/3$ must be done carefully.

For the case in which the seesaw expression (1.1) is in a good approximation, 
i.e., except for $b_f e^{i\beta_f} \simeq -1/3$, we can obtain 
simpler expressions of $m_i^f$ :
$$
m_1^f = z_1^2 \left|\frac{c^f_0}{c^f_1}\right| \frac{\kappa}{\lambda} m_0 \ \ , 
\eqno(3.1)
$$
$$
m_2^f = z_2^2 \left|\frac{c^f_1}{c^f_2}\right| \frac{\kappa}{\lambda} m_0 \ \ , 
\eqno(3.2)
$$
$$
m_3^f = z_3^2 \left|\frac{c^f_2}{c^f_3}\right| \frac{\kappa}{\lambda} m_0 \ \ , 
\eqno(3.3)
$$
where the functions $c^f_n \equiv c_n (b_f, \beta_f) \ (n = 1, 2, 3)$ 
are defined by 
$$
c^f_n \equiv c_n (b_f, \beta_f) = n + \frac{1}{b_f e^{i\beta_f}} \ \ . 
\eqno(3.4)
$$
Although the expressions (3.1) -- (3.3) are not valid for the cases 
$b_f e^{i\beta_f}=-1, -1/2$ and $-1/3$, these are still very 
useful for the case 
$\beta_f \neq 0$. 

More precise expressions for arbitrary values of $b_f$ and $\beta_f$ are 
obtained as follows. 
For the case of $\lambda \gg \kappa \gg 1$, 
by expanding the eigenvalues $m_i^f$ ($i=1,2,3$) 
of the mass matrix (2.1) in $\kappa/\lambda$,  
we obtain the following expressions of $m_i^f$:
$$
\left(\dfrac{m_1^f}{m_0}\right)^2 = \dfrac{2\sigma^2}{\rho^2 f(b,\beta)}
\left( 1+\sqrt{1-\dfrac{4\sigma^2}{\rho^4} \dfrac{g(b,\beta)}{f^2(b,\beta)}}
\, \right)^{-1}
\left(\dfrac{\kappa}{\lambda}\right)^2 
+O\left(\dfrac{\kappa^4}{\lambda^4}\right) \ , \eqno(3.5)
$$
$$
\left(\dfrac{m_2^f}{m_0}\right)^2  = \dfrac{\rho^2 f(b,\beta)}{g(b,\beta)}
\left( 1+\sqrt{1-\dfrac{4\sigma^2}{\rho^4} \dfrac{g(b,\beta)}{f^2(b,\beta)}}\, 
\right)
\left( 1+\sqrt{1+4\rho^2 \dfrac{f(b,\beta)h(b,\beta)}{g^2(b,\beta)}}
\, \right)^{-1}
\left(\dfrac{\kappa}{\lambda}\right)^2 
$$
$$
+O\left(\dfrac{\kappa^4}{\lambda^4}\right) \ , \eqno(3.6)
$$
$$
\left(\dfrac{m_3^f}{m_0}\right)^2  = 3g(b,\beta)
\left[
\left(\dfrac{\kappa}{\lambda}\right)^2+6h(b,\beta)
\left( 1+\sqrt{1+4\rho^2 \dfrac{f(b,\beta)h(b,\beta)}{g^2(b,\beta)}}
\, \right)^{-1}
\right]^{-1} 
 \left(\dfrac{\kappa}{\lambda}\right)^2 
$$
$$
+O\left(\dfrac{\kappa^4}{\lambda^4}\right) \ , \eqno(3.7)
$$
where
$$
f(b,\beta)=(1+b)^2  - 2(1+2b) \dfrac{\sigma}{\rho^2} 
-4b\left(1-2\dfrac{\sigma}{\rho^2}\right)\sin^2\dfrac{\beta}{2}
\ , \eqno(3.8)
$$
$$
g(b,\beta)=(1+2b)^2  - 2(1+b)(1+3b)\rho -8b(1-2\rho)\sin^2\dfrac{\beta}{2}
\ , \eqno(3.9)
$$
$$
h(b,\beta)=(1+3b)^2  -12b\sin^2\dfrac{\beta}{2}
\ , \eqno(3.10)
$$
$$
\rho=z_1^2 z_2^2 + z_2^2 z_3^2 + z_3^2 z_1^2 \ , \eqno(3.11)
$$
$$
\sigma=z_1^2 z_2^2 z_3^2  \ , \eqno(3.12)
$$
and for simplicity we have denoted $b_f$ and $\beta_f$ as $b$ and 
$\beta$.

Now let us apply the results (3.5)--(3.7) to the quark masses.
Our interest is in the cases $b_f \simeq -1/3$ and $b_f \simeq -1$ 
whose values are favorable to the fitting of the up- and down-quark 
masses, respectively. 
The explicit expressions are as follows:
$$
m_u \simeq \dfrac{3\sigma}{2\rho}\left( 
1+\dfrac{3\sigma}{4\rho^2} -\dfrac{3}{2}\varepsilon_u
\right) \dfrac{\kappa}{\lambda_U} m_0 
\simeq \dfrac{3m_e}{2m_\tau} \dfrac{\kappa}{\lambda_U} m_0 \ ,
\eqno(3.13)
$$ 
$$
m_c \simeq 2\rho \left[ 1-\dfrac{3\sigma}{4\rho^2}
-\dfrac{9}{2}\left(1-\dfrac{8}{3} \rho\right)\varepsilon_u  \right]
\dfrac{\kappa}{\lambda_U} m_0 \simeq 
2 \dfrac{m_\mu}{m_\tau} \dfrac{\kappa}{\lambda_U} m_0 \ , \eqno(3.14)
$$
$$
m_t \simeq \dfrac{1}{\sqrt{3}} \dfrac{1}{\sqrt{1 + 
27 \varepsilon_u^2 \lambda_U^2/\kappa^2}}
m_0 \simeq \dfrac{1}{\sqrt{3}}m_0 \ , \eqno(3.15)
$$
$$
m_d \simeq \dfrac{\sigma}{2|\sin(\beta_d/2)|\rho}
\left(1+\dfrac{1}{2}\varepsilon_d
\right) \dfrac{\kappa}{\lambda_D} m_0 \simeq 
\dfrac{1}{2|\sin(\beta_d/2)|}\dfrac{m_e}{m_\tau}
\dfrac{\kappa}{\lambda_D} m_0 \ , 
\eqno(3.16)
$$
$$
m_s \simeq 2\left(1+\dfrac{3}{2}\varepsilon_d 
-2\sin^2\dfrac{\beta_d}{2}\right)
\left|\sin\dfrac{\beta_d}{2}\right|\, \rho \dfrac{\kappa}{\lambda_D} 
m_0 \simeq 
2\left|\sin\dfrac{\beta_d}{2}\right| \dfrac{m_\mu}{m_\tau}
\dfrac{\kappa}{\lambda_D} m_0 
\ , \eqno(3.17)
$$
$$
m_b \simeq \dfrac{1}{2}\left(1-\dfrac{1}{2}\varepsilon_d 
+ \dfrac{5}{2}\sin^2 \dfrac{\beta_d}{2}\right) 
\dfrac{\kappa}{\lambda_D} m_0 
\simeq \dfrac{1}{2} \dfrac{\kappa}{\lambda_D} m_0 \ , \eqno(3.18)
$$
where the small parameters $\varepsilon_u$ and $\varepsilon_d$ 
are defined by
$$
\begin{array}{l}
b_u = -\dfrac{1}{3} + \varepsilon_u \ , \\
b_d = -1 + \varepsilon_d \ . 
\end{array} \eqno(3.19)
$$
Here, we have taken $\beta_u = 0$, because top-quark enhancement is 
caused only for the case of $\beta_u = 0$ (see Fig.~1). 
For down-quark masses, we have shown only the expressions for 
$b_d \simeq -1$ and $1 \gg \sin \beta_d \neq 0$, 
because from the numerical study in Ref.~3), we know 
that the observed down-quark mass spectrum is in favor of 
$b_d \simeq -1$ and $|\beta_d| \simeq 20^{\circ}$. 

The expressions (3.13) -- (3.19) lead to the following relations 
which are almost independent of the parameters $\kappa/\lambda$ 
($\lambda\equiv \lambda_U=\lambda_D$), 
$\varepsilon_u$, $\varepsilon_d$ and $\beta_d$: 
$$
\dfrac{m_u}{m_c} \simeq \dfrac{3}{4} \dfrac{m_e}{m_\mu} \ , 
\eqno(3.20)
$$
$$
\dfrac{m_c}{m_b} \simeq 4 \dfrac{m_\mu}{m_\tau}\ , 
\eqno(3.21)
$$
$$
\dfrac{m_d m_s}{m_b^2} \simeq 4 \dfrac{m_e m_\mu}{m_\tau^2} \ , 
\eqno(3.22)
$$
$$
\dfrac{m_u}{m_d} \simeq 3 \dfrac{m_s}{m_c} \simeq \dfrac{3}{4} 
\dfrac{m_d}{m_b} \dfrac{m_\tau}{m_\mu} \simeq 
3\left|\sin\dfrac{ \beta_d}{2}\right| \ . 
\eqno(3.23)
$$
The expressions (3.20) and (3.23) have already been given in 
Refs. 12) and 3), respectively. 
However, note that these relations are valid only for 
small value of $\varepsilon_u$ and $\varepsilon_d$, and not for general 
value of $b_f$.

In the limit of unbroken SU(2)$_L \times $SU(2)$_R$, i.e., 
$m_L=m_R=0$, heavy fermion masses $m_{F'_i}$ are given by 
$$
\begin{array}{ll}
m_{F'_1}=m_{F'_2}= \lambda_F m_0\ , \\[.2in]
m_{F'_3}=\sqrt{1+6 b_f \cos\beta_f +9b_f^2} \lambda_F m_0 \ ,
\end{array} \eqno(3.24)
$$ 
where $F'_i$ are mass-eigenstates for the mass matrix 
$M_F = m_0 \lambda_F O_f$. 
As seen from (3.24), the minimum condition of the sum of 
the up-heavy-quark masses leads to $\beta_u = 0$ and $b_u = -1/3$. 
Therefore, the ansatz ``maximal top-quark-mass enhancement" can 
be replaced by another expression that the parameters $(b_u, \beta_u)$ 
are fixed such that the sum of the up-heavy-quark masses becomes a minimum. 

For the case of $Z\neq 0$, the heavy fermion masses are given by 
$$
m_4^e \simeq m_5^e \simeq m_6^e \simeq \lambda_E m_0 \ , \eqno(3.25) 
$$
$$
m_4^u \simeq \dfrac{1}{\sqrt{3}} \kappa m_0, \ \ \ 
m_5^u \simeq m_6^u \simeq \lambda_U m_0 \ , 
\eqno(3.26)
$$
$$
m_4^d \simeq m_5^d \simeq \lambda_D m_0, \ \ \ 
m_6^d \simeq 2 \sqrt{1 + 3 \sin^2(\beta_d/2)} \lambda_D m_0 \ , \eqno(3.27)
$$
where the numbering of $m_i^f$ has been defined as 
$m_4^f \leq m_5^f \leq m_6^f$ in the mass eigenstates 
$F'_i$ $(i = 1, 2, 3)$. 
Note that only the fourth up-quark 
$u_4$ ($\equiv U'_3$) is remarkably light compared with 
other heavy fermions.  
The enhancement of the top-quark $u_3$ ($\equiv t$) is caused 
at the cost of the lightening of $U'_3$. Since the 
mass ratio $m_4^u/m_3^u$ is given by 
$$
m_4^u/m_t \simeq \kappa \eqno(3.28)
$$
and $\kappa$ is of the order of $m(W_R)/m(W_L)$, we can expect 
the observation of the fourth up-quark $u_4$ at an energy 
scale at which we can observe the right-handed weak bosons 
$W_R$. 


\vglue.2in
\centerline{\bf \S 4. General expression of family-mixing matrix}

\vglue.05in

We diagonalize the $6\times 6$ mass matrix $M$, (2.1), 
by the following two steps. 
As the first step, we transform the mass matrix $M$  into 
$$
M'=\left(
\begin{array}{cc}
M'_{11} & 0 \\
0 & M'_{22} 
\end{array}\right) 
\equiv \left(
\begin{array}{cc}
M_{f} & 0 \\
0 & M'_{F} 
\end{array}\right) 
\ . \eqno(4.1)
$$
At the second step, we diagonalize 
the $3\times 3$ matrix $M_f\equiv M'_{11}$ with $P_L^f=P_R^f={\bf 1}$ 
(which we denote as $\widetilde{M}_f$) 
by two unitary matrices $U_L^f$ and $U_R^f$ as follows:
$$
 U_L^f \widetilde{M}_f U_R^{f\dagger} = D_f \ , \eqno(4.2)
$$
where $D_f={\rm diag}(m_1^f, m_2^f, m_3^f)$.
Then, the CKM matrix $V$ is given by 
$$
V \simeq U_L^u P U_L^{d\dagger} \ , \eqno(4.3)
$$
where the phase matrix $P$ is defined by (2.11) and 
terms with the order of $\lambda^{-2}$ which come from the 
$f$-$F$ mixing have been neglected.

We denote the unitary matrix $U_L^f$ as
\renewcommand{\arraystretch}{2}
$$
U_L^f \simeq \left( 
\begin{array}{ccc} 
\displaystyle 
1-\varepsilon_1^f & 
\displaystyle (1-\varepsilon^f_{12})p_f \dfrac{z_1}{z_2} 
& \displaystyle (1-\varepsilon_{13}^f)p_f \dfrac{z_1}{z_3} \\
\displaystyle -(1-\varepsilon_{21}^f)p_f^* \dfrac{z_1}{z_2} 
& 1-\varepsilon_2^f & 
\displaystyle (1-\varepsilon_{23}^f)q_f \dfrac{z_2}{z_3} \\
 \displaystyle -(1-\varepsilon_{31}^f)q^*_f \dfrac{z_1}{z_3} 
& \displaystyle -(1-\varepsilon_{32}^f)q^*_f \dfrac{z_2}{z_3} 
& 1-\varepsilon_3^f
\end{array} \right) \ , \eqno(4.4)
$$
where the functions $p_f \equiv p(b_f, \beta_f)$ and 
$q_f \equiv q(b_f, \beta_f)$ are given by 
$$
p_f \equiv p(b_f, \beta_f) = \frac{b_f e^{i\beta_f}}{1 + b_f e^{i\beta_f}} 
= \frac{1}{c^f_1} \ \ , 
\eqno(4.5)
$$
$$
q_f \equiv q(b_f, \beta_f) = \frac{b_f e^{i\beta_f}}{1 + 2b_f e^{i\beta_f}} = 
\frac{1}{c_2^f} \ \ , 
\eqno(4.6)
$$
with the relation 
$$
c_2^f - c_1^f = 1 \ \ . 
\eqno(4.7)
$$

The next leading terms $\varepsilon_i^f$ and $\varepsilon_{ij}^f$ are 
obtained by putting the expression (4.4) into the unitary 
condition $U_L^f U_L^{f\dagger} = {\bf 1}$ and the 
diagonalization condition $U_L^f \widetilde{M}_f \widetilde{M}^\dagger_f 
U_L^{f\dagger} = D_f^2$:
$$
\varepsilon_1^f = \frac{1}{2}\left(\frac{z_1}{z_2}\right)^2 
\frac{1}{|c_1^f|^2} \ \ , 
\eqno(4.8)
$$
$$
\varepsilon_3^f = \frac{1}{2}\left(\frac{z_2}{z_3}\right)^2 
\frac{1}{|c_2^f|^2} \ \ , 
\eqno(4.9)
$$
$$
\varepsilon_2^f = \varepsilon_1^f + \varepsilon_3^f \ \ , 
\eqno(4.10)
$$
$$
\varepsilon_{12}^f = (3 - 2 c_1^{f2}) \varepsilon_1^f \ \ , 
\eqno(4.11)
$$
$$
\varepsilon_{21}^f = (3 - 2 c_1^{f*2}) \varepsilon_1^f - (c_1^{f*} + c_2^{f*}) 
\varepsilon_3^f \ \ , 
\eqno(4.12)
$$
$$
\varepsilon_{23}^f = (c_1^f + c_2^f) \varepsilon_1^f + 
(3 - 2 c_2^{f2}) \varepsilon_3^f \ \ , 
\eqno(4.13)
$$
$$
\varepsilon_{32}^f = (3 - 2 c_2^{f*2}) \varepsilon_3^f \ \ , 
\eqno(4.14)
$$
$$
\varepsilon_{13}^f = (3 - 2 c_1^f)\varepsilon_1^f \ \ , 
\eqno(4.15)
$$
$$
\varepsilon_{31}^f = (3 + 2 c_2^{f*}) \varepsilon_3^f \ \ . 
\eqno(4.16)
$$

The expression (4.4) is valid as far as we can regard $\varepsilon_1^f$ and 
$\varepsilon_3^f$ as $\varepsilon_1^f \ll 1$ and $\varepsilon_3^f \ll 1$, 
i.e., 
$$
|c_1^f|^2 = \left(\frac{1}{b_f} + 1\right)^2 
-\frac{4}{b_f} \sin^2\frac{\beta_f}{2}
\gg \frac{1}{2} \left(\frac{z_1}{z_2}\right)^2 = 0.0024 \ \ , 
\eqno(4.17)
$$
$$
|c_2^f|^2 = \left(\frac{1}{b_f} + 2\right)^2 
-\frac{8}{b_f} \sin^2\frac{\beta_f}{2} 
\gg \frac{1}{2} \left(\frac{z_2}{z_3} \right)^2 = 0.030 \ \ . 
\eqno(4.18)
$$
Therefore, for the cases $b_f = -1$ and $b_f = -1/2$, the expression (4.4) is 
valid only for the cases 
$$
\left|\sin\frac{\beta_f}{2} \right| \gg \frac{1}{2\sqrt{2}}\frac{z_1}{z_2} 
= 0.025 \  , \ \ \ 
(|\beta_f| \gg 2.8^\circ) \ \ , 
\eqno(4.19)
$$
and 
$$
\left|\sin\frac{\beta_f}{2} \right| \gg \frac{1}{4\sqrt{2}}\frac{z_2}{z_3} 
= 0.043 \ , \ \ \ 
(|\beta_f| \gg 4.9^\circ) \ \ , 
\eqno(4.20)
$$
respectively. 
For down-quark sector, we know that $b_d \simeq -1$ and 
$|\beta_d|\simeq 20^\circ$ from the phenomenological study${}^{3)}$ of 
the quark mass ratios. 
The value $|\beta_d| \simeq 20^\circ$ satisfies the condition (4.19), 
so that we can use the expression (4.4) for the down-quark sector. 
The expression (4.4) is not valid for the cases $b_f=-1$ and $b_f = -1/2$ 
with $\beta_f=0$, which do not satisfy the conditions (4.19) and (4.20).
The expressions for these cases are given in Appendix A.


\vglue.2in
\centerline{\bf \S 5. CKM matrix elements}

\vglue.05in

The CKM matrix elements $V_{ij}$ are given by (4.3).
Without losing  generality, we can take 
$$
P={\rm diag}(1,e^{i\delta_2}, e^{i\delta_3}) \ . 
\eqno(5.1)
$$

For the up-quark sector, we put an ansatz ``maximal top-quark-mass 
enhancement", i.e., we assume that $b_u = -1/3$ and $\beta_u = 0$. 
Then, from (4.5) -- (4.16), we obtain 
$$
p_u = p(-\frac{1}{3}, 0) = -\frac{1}{2} \ \ , \ \ \ q_u = 
q(-\frac{1}{3}, 0) 
= - 1 \ \ , 
\eqno(5.2)
$$
and
$$
\varepsilon_3^u = \frac{1}{2} \left(\frac{z_2}{z_3}\right)^2 =0.030
\gg 
\varepsilon_1^u = \frac{1}{8} \left(\frac{z_1}{z_2}\right)^2
=0.0006 \ \ , 
\eqno(5.3)
$$
$$
\varepsilon_{12}^u \simeq \varepsilon_{13}^u 
\simeq 0 \ , \ \ \varepsilon_{21}^u 
\simeq 
3 \varepsilon_3^u \ , \ \ \varepsilon_{23}^u \simeq 
\varepsilon_{31}^u \simeq 
\varepsilon_{32}^u \simeq \varepsilon_3^u \ ,
\eqno(5.4)
$$
with $\varepsilon_1^u \simeq 0$. Therefore, the unitary matrix 
$U_L^u$ is given by
\renewcommand{\arraystretch}{2}
$$
U_L^u \simeq \left(
\begin{array}{ccc}
1 & -\dfrac{1}{2}\dfrac{z_1}{z_2} & -\dfrac{1}{2}\dfrac{z_1}{z_3} \\
\dfrac{1}{2}\dfrac{z_1}{z_2}(1 - 3\varepsilon_3^u) & 
1 - \varepsilon_3^u & -\dfrac{z_2}{z_3}(1 - \varepsilon_3^u) \\
\dfrac{z_1}{z_3}(1 - \varepsilon_3^u) & 
\dfrac{z_2}{z_3}(1 - \varepsilon_3^u) & 
1 - \varepsilon_3^u  \\
\end{array} \right) \ \ . 
\eqno(5.5)
$$
\renewcommand{\arraystretch}{1}

For the down-quark sector, for a time, 
we use the general expression (4.4) without assuming $b_d \simeq -1$ and 
$\beta_d^2 \ll 1$.

First, by neglecting $\varepsilon_i^f$ and $\varepsilon_{ij}^f$ terms,
let us give rough estimates of $V_{ij}$:
$$
V_{us} \simeq -\frac{1}{2} \frac{z_1}{z_2} \frac{1}{c_1^d} e^{i \delta_2} 
(2 e^{-i \delta_2} + c_1^d) \ \ , 
\eqno(5.6)
$$
$$
V_{cb} \simeq -\frac{z_2}{z_3} \frac{1}{c_2^d} e^{i \delta_2} (1 + c_2^d 
e^{i (\delta_3 - \delta_2)}) \ \ , 
\eqno(5.7)
$$
$$
V_{ub} \simeq -\frac{1}{2} \frac{z_1}{z_3} \frac{1}{c_2^d} e^{i \delta_2} 
\left(2e^{-i \delta_2} - 1 + c_2^d e^{i (\delta_3 - \delta_2)}\right) \ \ , 
\eqno(5.8)
$$
$$
V_{td} \simeq \frac{z_1}{z_3} \frac{1}{c_1^{d*}} e^{i \delta_2} (c_1^{d*} 
e^{-i \delta_2} + 1 + e^{i (\delta_3 - \delta_2)}) \ \ . 
\eqno(5.9)
$$

In, (5.1), we have taken $\delta_1 =0$ without losing generality.  
We suppose that $\delta_2$ is also $\delta_2\simeq 0$. 
For  $\delta_2^2\ll \sin^2(\beta_d/2)$ and $\delta_2^2\ll\varepsilon_d^2$  
[$b_d\equiv -(1-\varepsilon_d)$], the relations (5.6) -- (5.9) lead to 
$$
V_{us} \simeq -\frac{1}{2} \frac{z_1}{z_2} \frac{c_3^d}{c_1^d}  \ \ , 
\eqno(5.10)
$$
$$
V_{cb} \simeq -\frac{z_2}{z_3} \frac{1}{c_2^d}  \left( c_2^d 
+ e^{-i (\delta_3 - \delta_2)}\right) e^{i (\delta_3 - \delta_2)} \ \ , 
\eqno(5.11)
$$
$$
V_{ub} \simeq -\frac{1}{2} \frac{z_1}{z_3} \frac{1}{c_2^d} \left( c_2^d 
+ e^{-i (\delta_3 - \delta_2)}\right) e^{i (\delta_3 - \delta_2)} \ \ , 
\eqno(5.12)
$$
$$
V_{td}^* \simeq \frac{z_1}{z_3} \frac{1}{c_1^{d*}}  \left( c_2^d 
+ e^{-i (\delta_3 - \delta_2)}\right) \ \ . 
\eqno(5.13)
$$
We can readily obtain an approximate relation 
$$
\left|\frac{V_{ub}}{V_{cb}} \right| \simeq \frac{1}{2} \frac{z_1}{z_2} 
= \frac{1}{2} \sqrt{\frac{m_e}{m_\mu}} = 0.035 \ \ , 
\eqno(5.14)
$$
which is valid for arbitrary values of the parameters $b_d$, 
$\beta_d$ and $(\delta_3 - \delta_2)$. 
However, the prediction (5.14) is somewhat small compared with 
the observed value${}^{7)}$ $|V_{ub}/V_{cb}| = 0.08 \pm 0.02$. 
This discrepancy can be corrected by taking 
the small terms $\varepsilon_i^f$ and $\varepsilon_{ij}^f$ into 
consideration (see Appendix B). 
We also obtain the relation 
$$
|V_{td}| \simeq 2 \left| \frac{c_2^d}{c_1^d} \right| \, |V_{ub}| 
\simeq 2 \sqrt{ \frac{m_\mu/m_\tau}{m_s/m_b} }
\frac{2|V_{us}|}{\sqrt{m_e/m_c}}\, |V_{ub}|  \ \ , 
\eqno(5.15)
$$
for arbitrary values of $b_d$, $\beta_d$ and $(\delta_3-\delta_2)$ 
by using the relations (3.2), (3.3), (5.10), (5.12) and (5.13).

{}From (5.10) and the relation 
$$
\frac{m_d}{m_s} \simeq \left(\frac{z_1}{z_2} \right)^2 
\frac{|c_0^d||c_2^d|}{|c_1^d|^2}  \ \ , 
\eqno(5.16)
$$
we can see that if we want to derive the well-known relation${}^{13)}$
$$
|V_{us}| \simeq \sqrt{m_d/m_s} \ \ , 
\eqno(5.17)
$$
we must impose a constraint 
$$
4 \simeq |c_3^d|^2/|c_0^d||c_2^d|  \ \ , 
\eqno(5.18)
$$
on the parameter $\beta_d e^{i\beta_d}$. 
The simplest one of the solutions of (5.18) is 
$b_d e^{i\beta_d} \simeq -1$, 
which yields reasonable down-quark mass ratios $m_d/m_s$ and $m_s/m_b$. 

Similarly, from (5.10), (5.11) and (5.13), we obtain the relation 
$$
\frac{|V_{td}|}{|V_{cb}| |V_{us}|} \simeq 
2 \left| \frac{c_2^d}{c_3^d} \right| \ \ . 
\eqno(5.19)
$$
Since $c_2^d\simeq 1$ and $c_3^d\simeq 2$ for $b_d e^{i\beta_d} \simeq -1$,
the ratio ($2|c_2^d/c_3^d|$) approximately takes one, 
so that we obtain the relation for the case of $b_d e^{i\beta_d} \simeq -1$, 
$$
\left| \frac{V_{td}}{V_{cb}}\right| \simeq |V_{us}| 
\simeq \sqrt{ \frac{m_d}{m_s} }  \ \ , 
\eqno(5.20)
$$
which is valid for arbitrary value of ($\delta_3 - \delta_2$). 
On the other hand, values of $|V_{td}|$, $|V_{cb}|$ and $|V_{us}|$ 
must be carefully estimated because those contain the small factor $c_1^d$ 
$$
c_1^d = \left[ \varepsilon_d -2i e^{-i\beta_d/2} \sin^2 
(\beta_d/2)\right]/(1-\varepsilon_d) \ \ , 
\eqno(5.21)
$$
which is sensitive to the values of $\varepsilon_d$ 
and $\beta_d$.

The rephasing invariant${}^{14)}$ $J$  is expressed 
in terms of $|V_{ij}|$ as follows:$^{15)}$
$$
J^2 = |V_{us}|^2 |V_{cb}|^2 |V_{ub}|^2 \left(
1+|V_{us}|^2-|V_{cb}|^2-\omega\right)
$$
$$
-\dfrac{1}{4}\left[ |V_{us}|^2 |V_{cb}|^2-\left(|V_{us}|^2+|V_{cb}|^2
\right) |V_{ub}|^2 +\left( 1-|V_{ub}|^2\right) \omega \right]^2 \ , 
\eqno(5.22)
$$
where
$$
\omega =|V_{cd}|^2-|V_{us}|^2=|V_{ts}|^2-|V_{cb}|^2
=|V_{ub}|^2-|V_{td}|^2
\ . \eqno(5.23)
$$
By using (5.20) and 
the observed fact $|V_{us}|^2\gg |V_{cb}|^2 \gg |V_{ub}|^2$, 
we obtain
$$
|J|\simeq \sqrt{
1-\dfrac{1}{4} \dfrac{|V_{ub} /V_{cb}|^2}{|V_{us}|^2 }}
|V_{us}|\, |V_{cb}|\, |V_{ub}|
$$
$$
\simeq \frac{1}{2}\sqrt{\frac{m_e}{m_\mu}\frac{m_d}{m_s}} 
\sqrt{ 1 -\frac{1}{4}\frac{m_e/m_\mu}{m_d/m_s} }\, 
|V_{cb}|^2 \ .
\eqno(5.24)
$$

For $\delta_2^2\ll \sin^2\beta_d +\varepsilon_d^2 \ll 1$, 
we obtain
$$
|V_{cb}| \simeq \frac{z_2}{z_3} \left| c_2^d + e^{-i(\delta_3 -
\delta_2)} \right| \ \ .
\eqno(5.25) 
$$
In order to explain the observed value${}^{7)}$ of $|V_{cb}|$ 
$$
|V_{cb}| = 0.041 \pm 0.003 \ \ , \eqno(5.26)
$$
the case $\delta_3-\delta_2\simeq 0$ is obviously ruled out 
because of $z_2/z_3=\sqrt{m_\mu /m_\tau}=0.244$ and $c_2^d\simeq 1$, 
and, rather, the case $\delta_3-\delta_2\simeq \pi$ is favorable 
to (5.26).
By putting  
$$
\delta_3 = \delta + \delta_2 + \pi \ \ , 
\eqno(5.27)
$$
we obtain 
$$
|V_{cb}| \simeq \frac{z_2}{z_3}
\sqrt{ \varepsilon_d^2 + (\sin\beta_d + \sin\delta )^2 } \ \ . 
\eqno(5.28)
$$
Similarly, for the case $|\delta_2|^2<|\delta|^2\ll 1$, we obtain
$$
|V_{td}| \simeq \frac{z_1}{z_3} \sqrt{ \frac{ \varepsilon_d^2 
+(\sin\beta_d +\sin\delta)^2 }{\varepsilon_d^2 
+\sin^2\beta_d} } \ \ .
\eqno(5.29)
$$

So far, we have not assumed $b_d=-1$. 
However, considering our parametrization $b_e=0$ and $b_u=-1/3$, 
it is likely that the value of $b_d$ is given not by $b_d\simeq -1$, 
but by a simple rational number $b_d=-1$. 
In Appendix B, we will show the more precise expressions of $|V_{ij}|$, 
in which we take the small terms $\varepsilon_i^f$ and 
$\varepsilon_{ij}^f$ given in (4.8) -- (4.16) into consideration, 
but we assume $b_d=-1$.

\vglue.2in
\centerline{\bf \S 6. Numerical results of the CKM matrix parameters}

\vglue.05in

Numerical results of $|V_{ij}|$ for $\delta_2 = \delta = 0$ have 
already been given in Ref.~3). Although the purpose of the 
present paper is not to give the numerical estimates, 
in order to complement the study of the previous section, 
in the present section, we shall give a numerical study of $|V_{ij}|$ 
without the restriction $\delta_2 = \delta = 0$.

As the numerical inputs, according to Ref.~3), 
we use $\kappa/\lambda=0.02$, $b_u=-1/3$, $\beta_u=0$, $b_d=-1$ and 
$\beta_d=18^\circ$, which are required for a reasonable 
fit with the observed quark masses.
Our interest is in the behavior of $|V_{ij}|$ versus the phase 
parameters $\delta_2$ and $\delta_3$ defined by (2.10) [(5.1)],
because in the previous study,$^{3)}$ the degree of freedom of the phases 
$(\delta_2, \delta_3)$ was not taken into consideration.
In Fig.~2, we illustrate the allowed regions of $(\delta_2, \delta_3)$ 
which give the observed values$^{7)}$ of $|V_{us}|$, $|V_{cb}|$ and 
$|V_{ub}|$:
$$
\begin{array}{l}
|V_{us}|= 0.2205 \pm 0.0018 \ , \\ 
|V_{cb}|=  0.041\pm 0.003 \ , \\
|V_{ub}/V_{cb}|= 0.08\pm 0.02 \ .
\end{array} \eqno(6.1)
$$
We have two allowed regions of $(\delta_2,\delta_3)$: 
we obtain the predictions 
$$
|V_{us}|= 0.2195 \ , \ \ \ |V_{cb}|= 0.0388  \ , 
\ \ \ |V_{ub}|=0.0028 \ ,
$$
$$
|V_{ub}/V_{cb}|= 0.072  \ , \ \ \ |V_{td}|=0.0105 \ , 
\ \ \ J= 1.8\times 10^{-5} \   ,
\eqno(6.2)
$$
for $(\delta_2,\delta_3)=(0^\circ , 174^\circ )$ and
$$
|V_{us}|= 0.2211 \ , \ \ \ |V_{cb}|= 0.0411 \ , \ \ \ |V_{ub}|=0.0027 \ ,
$$
$$
|V_{ub}/V_{cb}|= 0.065 \ , \ \ \ |V_{td}|=0.0092  \ , 
\ \ \ J= 2.4\times 10^{-5}  \   ,
\eqno(6.3)
$$
for $(\delta_2,\delta_3)=(-4^\circ , 152^\circ )$. 

In Fig.~3, we show the possible unitary-triangle shape of the present 
model on the $(\rho , \eta)$ plane, where $(\rho, \eta)$ are the 
Wolfenstein parameters$^{16)}$ defined by  
$V_{ub} \equiv |V_{us}||V_{cb}|(\rho - i \eta)$, 
$V_{us}=|V_{us}|$ and $V_{cb}=|V_{cb}|$.

The vertex $(\rho, \eta)$ moves on the circle which is denoted by 
the solid line in Fig.~3 according as the parameter $\delta_3$ varies from 
$0^\circ$ to $360^\circ$.
For reference, we have shown the constraints${}^{17)}$ 
from the observed values 
$|V_{ub}/V_{cb}|$, $\Delta m_{B_d}$ and $\varepsilon_K$.
Both triangles which correspond to the cases 
$(\delta_2, \delta_3) = (0^\circ, 174^\circ)$ and 
$(-4^\circ, 152^\circ)$ satisfy these constraints safely.

\vglue.2in
\centerline{\bf \S 7. Summary and discussion}

\vglue.05in

In conclusion, we have obtained the analytical expressions of 
the masses and mixings of the light fermions $f$ in the democratic seesaw 
mass matrix model (2.1). 

The fermion mass ratios are controlled by the parameters $b_f$, $\beta_f$ 
and $\kappa/\lambda_F$, as shown in Fig.~1. 
We have fixed the parameters $(z_1, z_2, z_3)$ by taking $b_e=0$ 
as given in (1.6).
The model can yield a large enhancement of top-quark mass, 
$m_t\gg m_b$ (keeping $m_u\sim m_d$), without taking hierarchically 
different values of mass matrix parameters in the up-quark sector. 
In the region of $(b_u\simeq -1/3, \beta_u\simeq 0)$ in which 
large top-quark-mass enhancement occurs, the mass relation (3.20), 
$m_u/m_c\simeq 3m_e/4m_\mu$, is valid almost independently of the 
parameter $\kappa/\lambda_F$ ($F=U$). 
The value of $\kappa/\lambda_U$ is fixed by the observed values of 
$m_c/m_t$. 
The observed down-quark mass values are in favor of $b_d\simeq -1$ 
with a small $\beta_d^2$ (but $\beta_d\neq 0$).
The mass relations (3.21)--(3.23) have been obtained for $b_d\simeq -1$ 
with a small $\beta_d^2$ and with $\lambda_D=\lambda_U$ 
($\equiv \lambda$).
Those relations are insensitive to value of $|\beta_d|$. 
The value of $|\beta_d|$ can be fixed by the observed mass ratio
$m_u/m_d$ (or $m_s/m_c$) as shown in (3.23).

As an application of the results, we have discussed the CKM matrix 
$V$. 
For the up-quark sector, we have assumed ^^ ^^ maximal top-quark-mass 
enhancement", i.e., $b_u = -1/3$ and $\beta_u =0$. 
We also suppose $\delta_2 \simeq 0$ in (5.1). 
Then, the relations (5.14) and (5.15) are valid almost independently 
of the parameters $b_d$, $\beta_d$ and ($\delta_3 - \delta_2$). 
The observed down-quark mass ratios is favorable to 
$b_d e^{i\beta_d} \simeq -1$. 
When we take $b_d e^{i\beta_d} \simeq -1$, the relations (5.17), 
(5.20) and (5.24) are valid almost independently of the value of 
$(\delta_3 - \delta_2)$. 
In order to fit $|V_{cb}|$ and $|V_{ub}|$ to the observed values, 
it is required that $\delta_3 - \delta_2 \simeq \pi$. 

Thus, in order to obtain a good fitting of the quark mass ratios and 
CKM mixings, we must take $b_u e^{i\beta_u} = -1/3$ 
and $b_d e^{i\beta_d} \simeq -1$ together with 
$b_e e^{i\beta_e} = 0$. 
The choice ($b_u = -1/3, \ \beta_u = 0$) is described by the ansatz 
of ^^ ^^ maximal top-quark-mass enhancement" or 
^^ ^^ minimal up-heavy-quark mass". 
However, the same ansatz cannot apply to the down-quark sector 
(it leads to a wrong solution $b_d e^{i\beta_d} = -1/3$). 

As an application of the fermion mass expressions (3.1)--(3.3), 
let us note the fermion mass ratio 
$$
r_f \equiv \dfrac{m_1^f m_3^f}{m_2^{f2}} = 
\dfrac{m_3^f/m_2^f}{m_2^f/m_1^f} = \dfrac{m_1^f/m_2^f}{m_2^f/m_3^f} \ \ , 
\eqno(7.1)
$$
which is expressed as 
$$
r_f = \left(\dfrac{z_1 z_3}{z_2^2}\right)^2 \dfrac{|c_0^f|}{|c_3^f|} 
\left(\dfrac{|c_2^f|}{|c_1^f|} \right)^3 \ \ . 
\eqno(7.2)
$$
Since 
$$
0 = \dfrac{\partial r_f}{\partial b_f} = -\dfrac{2|c_2^f|}{|c_3^f|^3 
|c_1^f|^5} b_f^2 (2b_f + \cos \beta_f)
$$
$$
\times (3b_f + 2\cos\beta_f + \sqrt{9 - 8\cos^2\beta_f})
(3b_f + 2\cos\beta_f - \sqrt{9 - 8\cos^2\beta_f}) \ \ , 
\eqno(7.3)
$$
the maximal points of $r_f$ are given by 
$$
b_f = -\frac{1}{3}(2\cos\beta_f - \sqrt{9 - 8\cos^2\beta_f}) \ \ , 
\eqno(7.4)
$$
and 
$$
b_f = -\frac{1}{3}(2\cos\beta_f + \sqrt{9 - 8\cos^2\beta_f}) \ \ . 
\eqno(7.5)
$$
For $\beta_f^2 \ll 1$, the former and the latter give 
$b_f \simeq -1/3$ and $b_f \simeq -1$, respectively. 
Although the expressions (3.1)--(3.3) are not valid for $b_f = -1/3$, 
$b_f = -1/2$ and $b_f = -1$ with $\beta_f = 0$,  we can see, 
from the numerical study of the $6 \times 6$ mass matrix, that the results 
(7.4) and (5.5) are valid even for $|\beta_f| \rightarrow 0$. 
Therefore, it is interesting to put an ansatz that the 
parameter $b_f$ takes its values such the ratio $r_f$ becomes maximal. 
Since $\partial r_f/\partial |\beta_f| < 0$, the ansatz leads to 
$|\beta_f| \rightarrow 0$. 
The solution ($b_f = -1/3, \ \beta_f = 0$) is favorable to the up-quark 
sector, but, for down-quark sector, the choice $\beta_f=0$ is not favorable. 
We will have to consider an additional reason for $\beta_f \neq 0$. 
If we accept such the additional condition $\beta_d\neq 0$ 
(but $\beta_d^2\ll 1$), we can obtain the desirable choice 
$b_d\simeq -1$.

In addition to the solutions (7.4) and (7.5), the remaining solutions of 
(7.3), $b_f = 0$ and $b_f= -\cos\beta_f/2\simeq -1/2$, are also interesting. 
The former $b_f=0$ 
corresponds to the case of the charged lepton sector. 
The latter $b_f\simeq -1/2$ 
is favorable to understanding a large neutrino 
mixing which has been suggested from the atmospheric neutrino 
data,$^{19)}$ as pointed out in Ref.~20) 
(also see (A.1) in Appendix A). 

Thus, although the ansatz for the mass ratio $r_f$ brings very interesting 
results, at present, we cannot find any plausible mechanism which justifies 
such the ansatz. 
Considering naively, it is strange that the parameter value of 
$b_f e^{i\beta_f}$ is controlled by the ratio $m_1^f m_3^f/m_2^{f2}$, 
because $b_f e^{i\beta_f}$ is a parameter in the heavy-fermion mass matrix 
$M_F$ which is generated at $\mu=m_0 \lambda_F$ ($\gg m_W$), 
while the masses $m_i^f$ ($i=1,2,3$) are generated at $\mu=m_0$ 
($\sim m_W$).
We consider that there is a fundamental law (mechanism) which 
controls the value of $b_f e^{i\beta_f}$ in $M_F$, and the law 
indirectly affects the mass ratio $r_f$, too. 
As a result, the points which yield $\partial r_f/\partial b_f=0$ can
correspond to the physical values of $b_f$ for quarks and leptons. 
However, it is a future task to clarify whether this scenario is 
true or not. 

The purpose of the present paper is to obtain analytical expressions 
of fermion masses and mixings for convenience of further investigating of the 
democratic seesaw mass matrix model. 
The model brings many new aspects beyond the conventional mass 
matrix models, and it seems to be worth while investigating 
the model furthermore.

\vglue.3in

\centerline{\large\bf Acknowledgments}

The authors would like to express their sincere thanks to Professors 
R.~Mohapatra, K.~Hagiwara and M.~Tanimoto for  their valuable comments 
on an earlier version of the present work, especially, to Prof.~M.~Tanimoto 
for pointing out the wrong sing of $J$ in the earlier version to the 
authors.
This work was supported by the Grant-in-Aid for Scientific Research, the 
Ministry of Education, Science and Culture, Japan (No.06640407 and 
No.08940386). 

\vspace{15mm}
\centerline{\large\bf Appendix A}
\vglue.1in

The expressions of $U_L^f$ for the cases $b_f = -1/2$ and $b_f = -1$ with 
$\beta_f = 0$ are given in Ref.~20). 
The results are as follows:
\renewcommand{\arraystretch}{2}
$$
U_L^f \simeq \left(\begin{array}{ccc}
1 & -\dfrac{z_1}{z_2} & -\dfrac{z_1}{z_3} \\
\dfrac{1}{\sqrt{2}}\left(\dfrac{z_1}{z_2} - \dfrac{z_1}{z_3}\right) & 
\dfrac{1}{\sqrt{2}} & -\dfrac{1}{\sqrt{2}} \\
\dfrac{1}{\sqrt{2}}\left(\dfrac{z_1}{z_2} + \dfrac{z_1}{z_3}\right) & 
\dfrac{1}{\sqrt{2}} & \dfrac{1}{\sqrt{2}} \\ 
\end{array} \right) \ \ , 
\eqno({\rm A}.1)
$$
\renewcommand{\arraystretch}{1}
for $b_f = -1/2$ and $\beta_f = 0$, and 
\renewcommand{\arraystretch}{2}
$$
U_L^f \simeq \left(\begin{array}{ccc}
\dfrac{1}{\sqrt{2}} & \dfrac{1}{\sqrt{2}} & 
\dfrac{1}{\sqrt{2}}\left(\dfrac{z_2}{z_3} - \dfrac{z_1}{z_3}\right) \\ 
-\dfrac{1}{\sqrt{2}} & \dfrac{1}{\sqrt{2}} & 
\dfrac{1}{\sqrt{2}}\left(\dfrac{z_2}{z_3} + \dfrac{z_1}{z_3}\right) \\ 
-\dfrac{z_1}{z_3} & -\dfrac{z_2}{z_3} & 1 \\
\end{array}
\right) \ \ , 
\eqno({\rm A}.2)
$$
\renewcommand{\arraystretch}{1}
for $b_f =-1$ and $\beta_f = 0$. 

\vspace{15mm}

\centerline{\large\bf Appendix B}
\vglue.1in

For $b_d = -1$, $\beta_d^2 \ll 1$ (but $\beta_d^2 \neq 0$), 
$\delta^2 \ll 1$ and $\delta_2^2 \ll 1$, 
from (4.4) with $b_f=-1$  and (5.5),  we obtain the following 
analytical expressions of $|V_{ij}|$: 
$$
|V_{us}| \simeq \frac{z_1}{2z_2} \frac{1}{|\sin(\beta_d/2)|} 
\left[1 - 3\varepsilon_1^d + \varepsilon_3^d + \sin\frac{\beta_d}{2} 
\sin\left(\frac{\beta_d}{2} - \delta_2\right) + \frac{1}{2} 
\sin^2 \frac{\beta_d}{2} 
\right] \ \ , 
\eqno({\rm B}.1)
$$
$$
|V_{cb}| \simeq \frac{z_2}{z_3} \frac{1 - \varepsilon_3^u - 
\frac{1}{2}\varepsilon_1^d - \frac{5}{2} \varepsilon_3^d}
{\sqrt{1 + 8\sin^2 (\beta_d/2)}} \left|\sin\beta_d + \sin\delta \right| 
(1 + \eta_{cb}) \ \ , 
\eqno({\rm B}.2)
$$
$$
|V_{ub}| \simeq \frac{z_1}{z_3} \frac{1 - \varepsilon_3^d}{\sqrt{1 + 8\sin^2 
(\beta_d/2)}} \left|\sin\beta_d + \sin\delta + 2\sin\delta_2 \right| 
(1 + \eta_{ub}) \ \ , 
\eqno({\rm B}.3)
$$
$$
|V_{td}| \simeq 2\frac{z_1}{z_3} \frac{|\sin\beta_d + \sin\delta|}{|\sin 
(\beta_d/2)|} (1 + \eta_{td}) \ \ , 
\eqno({\rm B}.4)
$$
where 
$$
\varepsilon_1^d = \frac{1}{8} \left(\frac{z_1}{z_2} \right)^2 
\frac{1}{\sin^2 (\beta_d/2)} \simeq \frac{1}{2} \frac{m_d}{m_s} 
\simeq \frac{1}{2}|V_{us}|^2 \ \ , 
\eqno({\rm B}.5)
$$
$$
\varepsilon_3^d = \frac{1}{2} \left(\frac{z_1}{z_3} \right)^2 
\frac{1}{\sqrt{1 + 8 \sin^2 
(\beta_d/2)}} \simeq \frac{1}{2} \sqrt{\frac{m_\mu}{m_\tau}} \ \ , 
\eqno({\rm B}.6)
$$
$$
\eta_{cb} = 2 \left[ \left(\sin^2 \frac{\beta_d}{2} + 
\sin^2 \frac{\delta}{2} \right)^2 - 2 \left(\varepsilon_1^d + 
\varepsilon_3^d \right) \sin^2 \frac{\beta_d}{2} + 
4 \varepsilon_3^d \sin^2 \frac{\delta}{2} + \frac{1}{4} 
\left(\varepsilon_1^d - \varepsilon_3^d \right)^2 \right] 
$$
$$
\times \left[ \left(1 - \varepsilon_3^u - \frac{1}{2} \varepsilon_1^d - 
\frac{5}{2} \varepsilon_3^d \right) \left(\sin\frac{\beta_d}{2} + 
\sin\frac{\delta}{2} \right) \right]^{-2} \ \ , 
\eqno({\rm B}.7)
$$
$$
\eta_{ub} = 2 \left\{ \left(\sin^2 \frac{\beta_d}{2} + \sin^2\frac{\delta}{2} 
\right)^2 + 2 \sin^2 \frac{\beta_d}{2} \left(\sin\frac{\delta}{2} + 
\sin\frac{\delta_2}{2} \right)^2 \right. 
$$
$$
- 8 \sin\frac{\beta_d}{2} \sin\frac{\delta}{2} \sin\frac{\delta_2}{2} 
\left(\sin\frac{\delta}{2} + \sin\frac{\delta_2}{2} \right) 
- 4 \left(\sin^2 \frac{\delta}{2} - \sin^2\frac{\delta_2}{2} \right) 
\sin^2 \frac{\delta_2}{2} 
$$
$$
\left.- 8 \varepsilon_b \left[-2 \sin \frac{\beta_d}{2} 
\left(\sin \frac{\beta_d}{2} - \sin\frac{\delta}{2} \right) + 
\left(\sin\frac{\delta}{2} + \sin\frac{\delta_2}{2} \right)^2 \right] 
\right\} 
$$
$$
\times \left[\left(1 - \varepsilon_3^d \right) \left(\sin \beta_d + 
\sin\delta + 2\sin\delta_2 \right) \right]^{-2} \ \ , 
\eqno({\rm B}.8)
$$
$$
\eta_{td} = 2 \left[ \left(\sin^2\frac{\beta_d}{2} + \sin^2 \frac{\delta}{2} 
\right)^2 + 4 \sin\frac{\beta_d}{2} \sin\frac{\delta}{2} 
\sin\frac{\delta_2}{2} \left(\sin\frac{\beta_d}{2} - \sin \frac{\delta}{2} 
- \sin\frac{\delta_2}{2} \right) \right. 
$$
$$
\left.- \varepsilon_1^d 
\left(\sin\frac{\beta_d}{2} + \sin \frac{\delta}{2} \right)^2 \right] 
\left[\left(1 - \varepsilon_1^d \right) \left(\sin\beta_d + \sin\delta 
\right) \right]^{-2} \ \ . 
\eqno({\rm B}.9)
$$


\newpage
\vspace{15mm}
\vglue.3in
\newcounter{0000}
\centerline{\large\bf References}
\begin{list}
{\arabic{0000}~)}{\usecounter{0000}
\labelwidth=0.8cm\labelsep=.1cm\setlength{\leftmargin=0.7cm}
{\rightmargin=.2cm}}

\item M.~Gell-Mann, P.~Rammond and R.~Slansky, in {\it Supergravity}, 
edited by P.~van Nieuwenhuizen and D.~Z.~Freedman (North-Holland, 
1979).
 
T.~Yanagida, in {\it Proc. Workshop of the Unified Theory and 
Baryon Number in the Universe}, edited by A.~Sawada and A.~Sugamoto 
(KEK, 1979).

R.~Mohapatra and G.~Senjanovic, Phys.~Rev.~Lett.~{\bf 44} (1980), 912.

\item Z.~G.~Berezhiani, Phys.~Lett.~{\bf 129B} (1983), 99; 
Phys.~Lett.~{\bf 150B} (1985), 177.

D.~Chang and R.~N.~Mohapatra, Phys.~Rev.~Lett.~{\bf 58} (1987), 1600.

A.~Davidson and K.~C.~Wali, Phys.~Rev.~Lett.~{\bf 59} (1987), 393.

S.~Rajpoot, Mod.~Phys.~Lett. {\bf A2} (1987), 307.
 
Phys.~Lett.~{\bf 191B} (1987), 122; Phys.~Rev.~{\bf D36} (1987), 1479.

K.~B.~Babu and R.~N.~Mohapatra, Phys.~Rev.~Lett.~{\bf 62}  (1989), 1079;
Phys.~Rev. {\bf D41} (1990), 1286. 

S.~Ranfone, Phys.~Rev.~{\bf D42} (1990), 3819.

A.~Davidson, S.~Ranfone and K.~C.~Wali, Phys.~Rev.~{\bf D41} (1990), 208.

I.~Sogami and T.~Shinohara, Prog.~Theor.~Phys.~{\bf 66} (1991), 1031;
Phys.~Rev.~{\bf D47} (1993), 2905.

Z.~G.~Berezhiani and R.~Rattazzi, Phys.~Lett.~{\bf B279} (1992), 124.

P.~Cho, Phys.~Rev.~{\bf D48} (1994), 5331.

A.~Davidson, L.~Michel, M.~L,~Sage and  K.~C.~Wali, Phys.~Rev.~{\bf D49} 
(1994), 1378.

W.~A.~Ponce, A.~Zepeda and R.~G.~Lozano, Phys.~Rev.~{\bf D49} (1994), 4954.
\item Y.~Koide and H.~Fusaoka, Z.~Phys.~{\bf C71} (1996), 459.
\item H.~Terazawa, University of Tokyo, Report No.~INS-Rep.-298 (1977) 
(unpublished); Genshikaku Kenkyu (INS, Univ.~of Tokyo) {\bf 26} 
(1982), 33.

Y.~Koide, Phys.~Rev. {\bf D49} (1994), 2638.
\item H.~Harari, H.~Haut and J.~Weyers, Phys.~Lett.~{\bf 78B} (1978), 459.

T.~Goldman, in {\it Gauge Theories, Massive Neutrinos and 
Proton Decays}, edited by A.~Perlumutter (Plenum Press, New York, 
1981), p.111.

T.~Goldman and G.~J.~Stephenson,~Jr., Phys.~Rev.~{\bf D24} (1981), 236.

Y.~Koide, Phys.~Rev.~Lett. {\bf 47} (1981), 1241; 
Phys.~Rev.~{\bf D28} (1983), 252; {\bf 39} (1989), 1391.

C.~Jarlskog, in {\it Proceedings of the International Symposium on 
Production and Decays of Heavy Hadrons}, Heidelberg, Germany, 1986
edited by K.~R.~Schubert and R. Waldi (DESY, Hamburg), 1986, p.331.

P.~Kaus, S.~Meshkov, Mod.~Phys.~Lett.~{\bf A3} (1988), 1251; 
Phys.~Rev.~{\bf D42} (1990), 1863.

L.~Lavoura, Phys.~Lett.~{\bf B228} (1989), 245.

M.~Tanimoto, Phys.~Rev.~{\bf D41} (1990), 1586.

H.~Fritzsch and J.~Plankl, Phys.~Lett.~{\bf B237} (1990), 451.
 
Y.~Nambu, in {\it Proceedings of the International Workshop on 
Electroweak Symmetry Breaking}, Hiroshima, Japan, (World 
Scientific, Singapore, 1992), p.1.
\item N.~Cabibbo, Phys.~Rev.~Lett.~{\bf 10} (1996), 531.

M.~Kobayashi and T.~Maskawa, Prog.~Theor.~Phys.~{\bf 49} (1973), 652.
\item Particle data group, R.~M.~Barnet {\it et al},  
Phys.~Rev.~{\bf D54} (1996), 1.
\item J.~C.~Pati and A.~Salam, Phys.~Rev. {\bf D10} (1974), 275.

R.~N.~Mohapatra and J.~C.~Pati, Phys.~Rev. {\bf D11} (1975), 366 and 258.

G.~Senjanovic and R.~N.~Mohapatra, Phys.~Rev. {\bf D12} (1975), 1502.
\item H.~Harari, H.~Haut and J.~Weyers, Phys.~Lett.~{\bf 78B} (1978), 459.
\item Y.~Koide, Phys.~Rev.~Lett. {\bf 47} (1981), 1241.
\item P.~Kaus, S.~Meshkov, Mod.~Phys.~Lett.~{\bf A3} (1988), 1251;
Phys.~Rev.~{\bf D42} (1990), 1863.

Y.~Nambu, in {\it Proceedings of the International Workshop on 
Electroweak Symmetry Breaking}, Hiroshima, Japan, (World 
Scientific, Singapore, 1992), p.1.
\item Y.~Koide, Mod.~Phys.~Lett.~{\bf A8} (1993), 2071.
\item S.~Weinberg, Ann.~N.Y.~Acad.~Sci.~{\bf 38} (1977), 1945.

H.~Fritszch, Phys.~Lett. {\bf 73B} (1978), 317; 
Nucl.~Phys.~{\bf B155} (1979), 189.

H.~Georgi and D.~V.~Nanopoulos, Nucl.~Phys.~{\bf B155} (1979), 52.
\item C.~Jarlskog, Phys.~Rev.~Lett.~{\bf 55} (1985)  1839.

O.~W.~Greenberg, Phys.~Rev. {\bf D32} (1985), 1841.

I.~Dunietz, O.~W.~Greenberg, and D.-d.~Wu, Phys.~Rev.~Lett. {\bf 55} 
 (1985), 2935.

C.~Hamzaoui and A.~Barroso, Phys.~Lett.~{\bf 154B} (1985), 202.

D.-d.~Wu, Phys.~Rev.~{\bf D33} (1986), 860.
\item C.~Hamzaoui, Phys.~Rev.~Lett.~{\bf 61} (1988), 35.

G.~C.~Branco and L.~Lavoura, Phys.~Lett.~{\bf B208} (1988), 123.
\item L.~Wolfenstein, Phys.~Rev.~{\bf D45} (1992), 4186.
\item For instance, see 

A.~Pich and J.~Prades, Phys.~Lett.~{\bf B346}
(1995), 342.
\item The ``family symmetry" is also called a ``horizontal symmetry":

K.~Akama and H.~Terazawa, Univ.~of Tokyo, Report No.~INS-Rep.-257 (1976) 
(unpublished). 

T.~Maehara and T.~Yanagida, Prog.~Theor.~Phys. {\bf 60} (1978), 822. 

F.~Wilczek and A.~Zee, Phys.~Rev.~Lett. {\bf 42} (1979), 421.

A.~Davidson, M.~Koca and K.~C.~Wali, Phys.~Rev. {\bf D20} (1979), 1195.

J.~Chakrabarti, Phys.~Rev. {\bf D20} (1979), 2411. 
\item Y.~Fukuda {\it et al}, Phys.~Lett. {\bf B335} (1994), 237.
\item Y.~Koide, US-95-07 (1995) (hep-ph/9508369);
 US-96-04 (1996) (hep-ph/9603376), to be publised in Mod.~Phys.~Lett. 
 (1996).
\end{list}

\newpage

\begin{center}
{\bf Figure Captions}
\end{center}

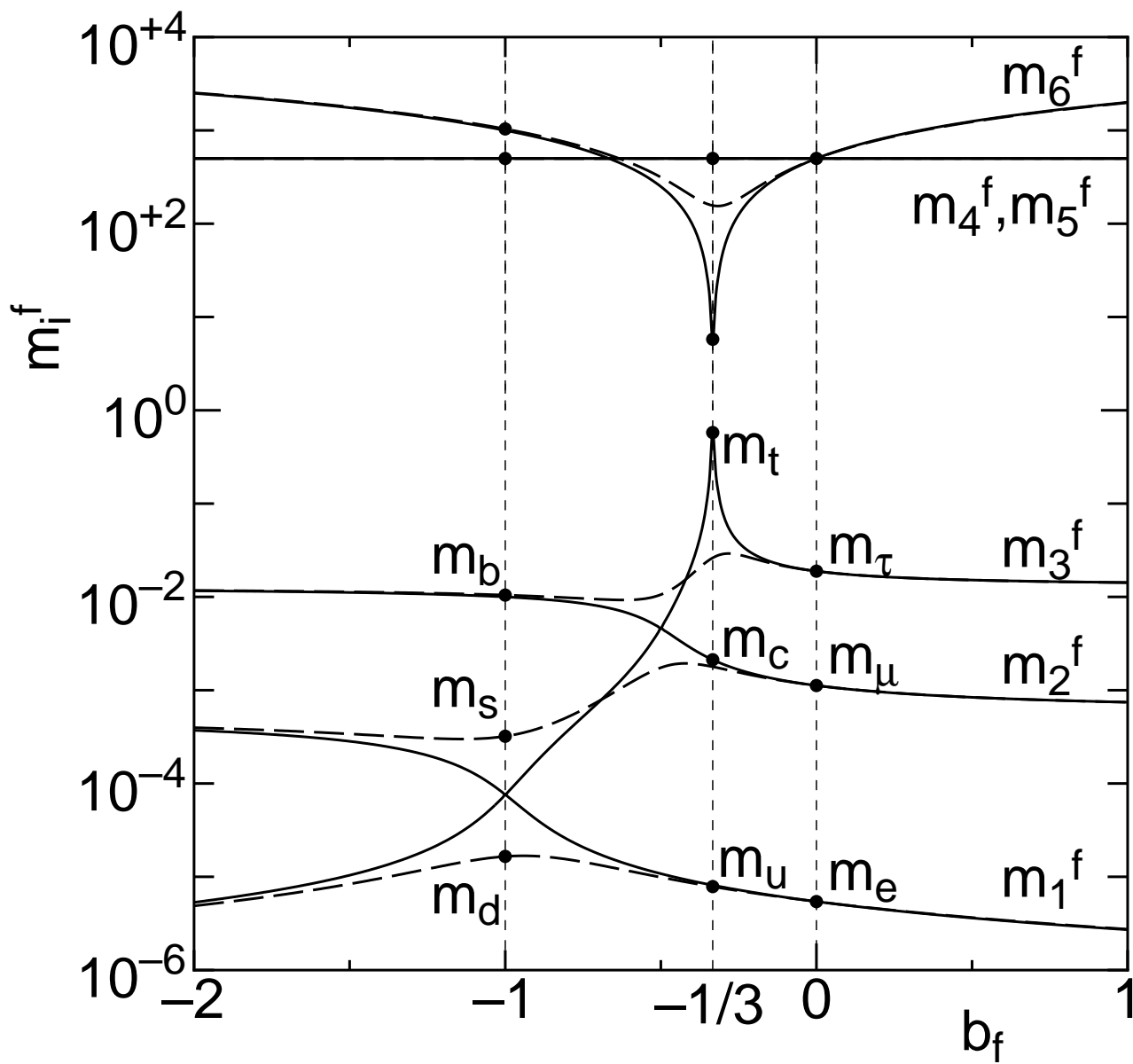
Fig.~1. Masses  $m_i^f$ ($i=1,\cdots,6$) versus $b_f$ for the case of 
$\kappa=10$ and  $\kappa/\lambda=0.02$.
The solid and broken lines denote for the cases of $\beta_f=0$ 
and $\beta_f=18^\circ$, respectively. 
At $b_f=0$, the charged lepton masses $m_e$, $m_\mu$ and $m_\tau$ 
have been used as input values for the parameters $z_i$.
For up- and down-quark sectors, the values $b_u=-1/3$ and $b_d=-1$ 
are chosen from the phenomenological study${}^{3)}$ of the observed 
quark masses.

\vglue.1in

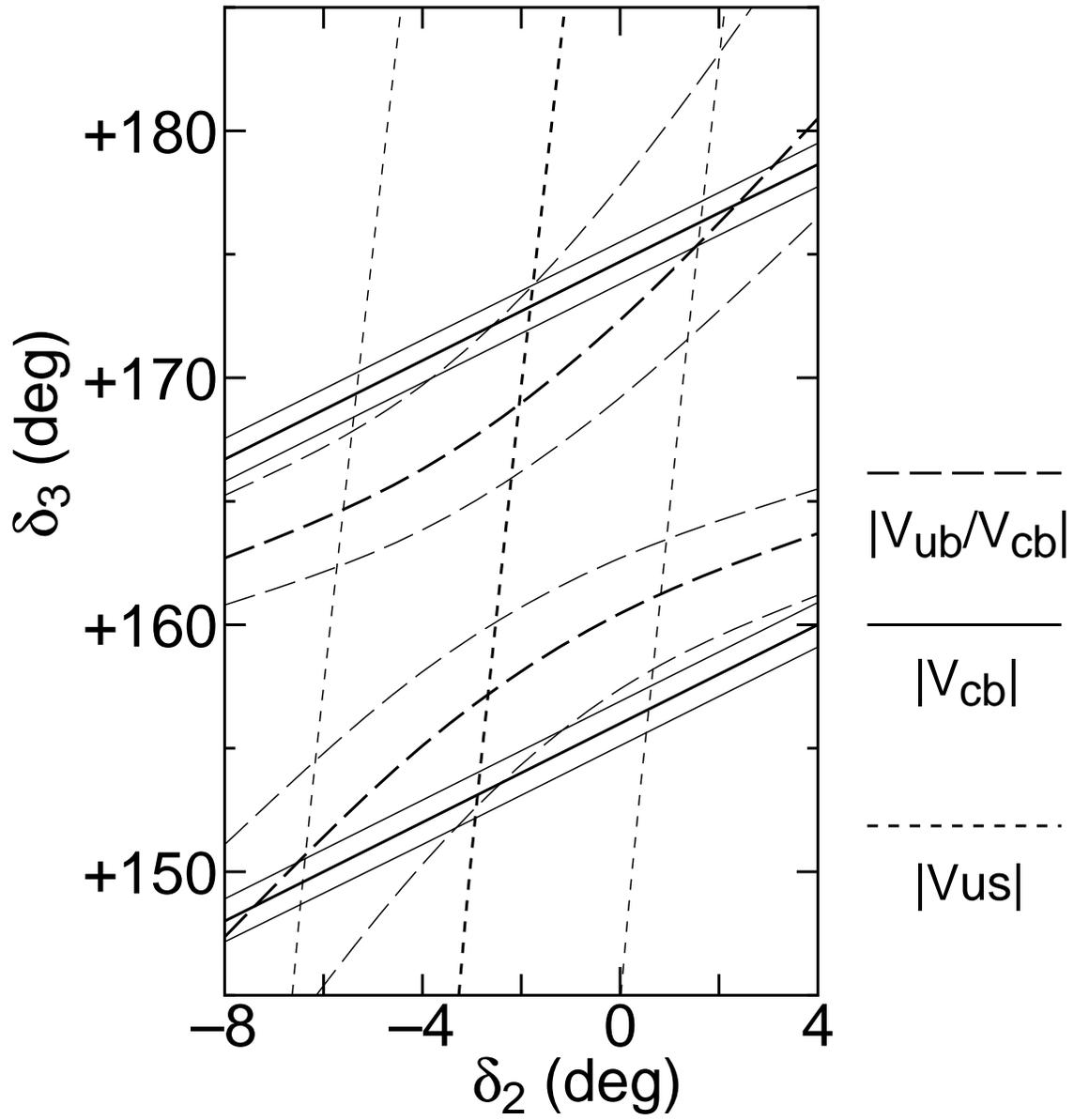
Fig.~2.  
Constraints on the phase parameters $(\delta_2, \delta_3)$ from 
the experimental values $|V_{us}|=0.2205\pm 0.0018$ (dotted lines), 
$|V_{cb}|=0.041\pm 0.003$ (solid lines) and 
$|V_{ub}/V_{cb}|=0.08\pm 0.02$ (dashed lines).
The hatched areas denote the allowed regions.

\vglue.1in

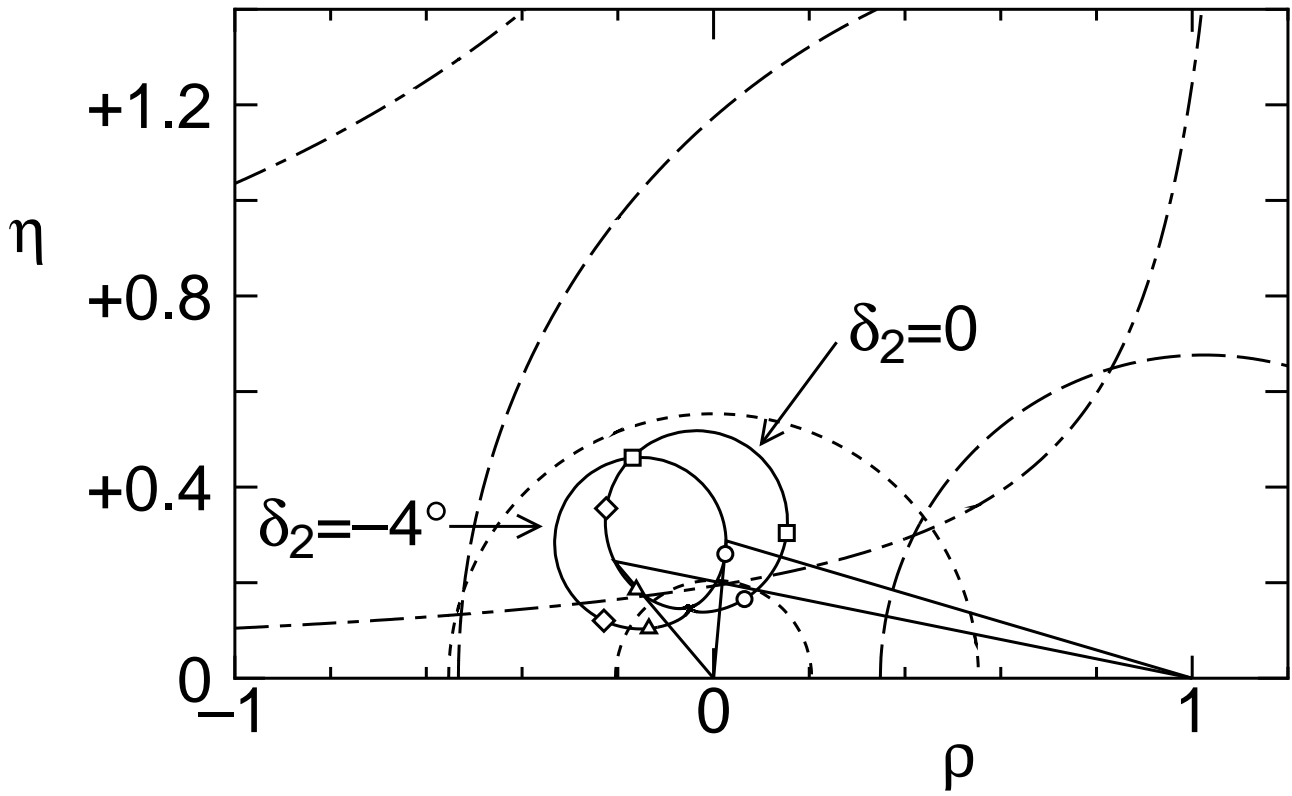
Fig.~3.  
Trajectories of the vertex $(\rho, \eta)$ of the unitary triangle 
for the cases  $\delta_2=0^\circ$ and $\delta_2=-4^\circ$. 
The points $\circ$, $\Box$, $\Diamond$ and $\triangle$ denote the vertex 
$(\rho, \eta)$ for $\delta_3=150^\circ$, $160^\circ$, $170^\circ$ 
and $180^\circ$, respectively.
The other parameters are fixed to $\kappa =10$, $\kappa/\lambda=0.02$, 
$b_u=-1/3$, $\beta_u=0$, 
$b_d=-1$, and $\beta_d=18^\circ$ from the observed quark mass ratios.
The solid, broken and dot-dashed lines denote constraints from 
$|V_{ub}/V_{cb}|$, $|\Delta m_{B_d}|$ and $\varepsilon_K$.
The two triangles correspond to the cases 
$(\delta_2, \delta_3)=(0^\circ, 174^\circ)$ and $(-4^\circ, 152^\circ)$, 
respectively.

\end{document}